\documentclass{aastex631}

\newcommand{\mathbi}[1]{\textbf{\textit{#1}}}

\shorttitle{Particle in the Current Sheet}
\shortauthors{Wu et al.}

\begin{document}

\title{Particle Acceleration and Transport in the Large-scale Current Sheet under an Erupting Magnetic Flux Rope}

\author[0009-0006-9249-8468]{H.~Wu}
\affiliation{Key Laboratory of Modern Astronomy and Astrophysics (Nanjing University), Ministry of Education, Nanjing 210023, People's Republic of China}

\author[0000-0002-9293-8439]{Y.~Guo}
\affiliation{Key Laboratory of Modern Astronomy and Astrophysics (Nanjing University), Ministry of Education, Nanjing 210023, People's Republic of China}

\author[0000-0003-3544-2733]{R.~Keppens}
\affiliation{Centre for mathematical Plasma Astrophysics, Department of Mathematics, KU Leuven, Celestijnenlaan 200B, B-3001 Leuven, Belgium}

\author[0000-0002-7153-4304]{C.~Xia}
\affiliation{School of Physics and Astronomy, Yunnan University, Kunming 650050, People's Republic of China}

\author[0000-0002-0197-470X]{Y.~Su}
\affiliation{Key Laboratory of Dark Matter and Space Astronomy, Purple Mountain Observatory, Chinese Academy of Sciences, Nanjing 210023, China}

\author[0000-0003-1034-5857]{X.~L.~Kong}
\affiliation{Institute of Space Sciences, Shandong University, Weihai, Shandong 264209, People's Republic of China}

\author[0000-0002-4978-4972]{M.~D.~Ding}
\affiliation{Key Laboratory of Modern Astronomy and Astrophysics (Nanjing University), Ministry of Education, Nanjing 210023, People's Republic of China}

\correspondingauthor{Y.~Guo}
\email{guoyang@nju.edu.cn}

\begin{abstract}

We investigate the acceleration and transport of electrons in the highly fine-structured current sheet that develops during magnetic flux rope (MFR) eruptions. Our work combines ultra-resolved MHD simulations of MFR eruption, with test-particle studies performed using the guiding center approximation. Our grid-adaptive, fully three-dimensional, high-resolution magnetohydrodynamic simulations model MFR eruptions that form complex current sheet topologies, serving as background electromagnetic fields for particle acceleration. Within the current sheet, tearing-mode instabilities give rise to mini flux ropes. Electrons become temporarily trapped within these elongated structures, undergoing acceleration and transport processes that significantly differ from those observed in two-dimensional or two-and-a-half-dimensional simulations. Our findings reveal that these fine-scale structures act as efficient particle accelerators, surpassing the acceleration efficiency of single X-line reconnection events, and are capable of energizing electrons to energies exceeding 100 keV. High-energy electrons accelerated in different mini flux ropes follow distinct trajectories due to spatially varying magnetic field connectivity, ultimately precipitating onto opposite sides of flare ribbons. Remarkably, double electron sources at the flare ribbons originate from different small flux rope acceleration regions, rather than from the same reconnecting field line as previously suggested. Distinct small flux ropes possess opposite magnetic helicity to accelerate electrons to source regions with different magnetic polarities, establishing a novel conjugate double source configuration. Furthermore, electrons escaping from the lower regions exhibit a broken power-law energy spectrum. This spectral break arises from electrons accelerated in disparate mini flux ropes, each exhibiting magnetic reconnection rates and acceleration efficiencies, which reflect the varying local reconnection conditions.
\end{abstract}


\section{Introduction} \label{sec:intro}

Solar flares are the most energetic events in the solar corona, characterized by the rapid release of magnetic energy. These events are pivotal in driving space weather phenomena, making their study essential for understanding solar activity and its impact on the solar system. High-energy observations, particularly hard X-ray (HXR) emissions, are crucial for probing the dynamics of solar flares, as they provide direct insights into electron acceleration processes. The HXR emission is interpreted as bremsstrahlung radiation from accelerated particles from the solar flare. Several space missions have been dedicated to observing these high-energy phenomena. The \emph{Reuven Ramaty High Energy Solar Spectroscopic Imager} \citep[\emph{RHESSI};][]{lin2002}, has significantly advanced our understanding of the nonthermal processes in solar flares. More recently, the \emph{Spectrometer/Telescope for Imaging X-rays} \citep[\emph{STIX};][]{krucker2020b} on the \emph{Solar Orbiter} (\emph{SolO}) mission and the \emph{Hard X-ray Imager} \citep[\emph{HXI};][]{zhang2019a, su2019, su2024} on the \emph{Advanced Space-based Solar Observatory} \citep[\emph{ASO-S};][]{gan2019,gan2023} continue to deliver new insights. HXR observations serve as effective indicators of electron acceleration in the solar corona, as evidenced by numerous observations of conjugate footpoint sources and possible loop-top sources \citep{kontar2011, chen2021, wu2023}.  In addition to high-energy emissions, the reconnection-driven particles also contribute to the Solar Energetic Particles (SEPs). These particles are detected by missions such as \emph{Solar Terrestrial Relations Observatory} (\emph{STEREO}), \emph{Geostationary Operational Environment Satellites} (\emph{GOES}), the \emph{WIND} spacecraft, and \emph{Parker Solar Probe} (\emph{PSP}).

Magnetic reconnection is a fundamental process in solar flares, responsible for transforming free, non-potential magnetic energy into the kinetic energy of particles \citep{fletcher2011,benz2017}. This phenomenon predominantly occurs in the current sheet, which is a long, stretched structure filled with electric currents, as described in the classical CSHKP model \citep{carmichael1964, sturrock1966, hirayama1974, kopp1976}. These events are often accompanied by the formation of two ribbons and a coronal mass ejection (CME). Despite its significance, the precise mechanisms of magnetic reconnection remain a topic of ongoing research. The two-dimensional, steady Sweet-Parker model proposes a thin and diffused current sheet, which only leads to slow energy release \citep{sweet1958, parker1957}. The Petschek model \citep{petschek1964} includes two slow-mode shocks that effectively accompany a thereby shortened current sheet to achieve faster energy release. Besides these two 2D and steady reconnection types, linear-stability theory suggests that exceeding a critical current sheet aspect ratio would lead to instability of current sheet at finite resistivity. The tearing instability generates the magnetic islands in a two-dimensional scenario \citep{oka2010a} and mini flux ropes in a three-dimensional situation \citep{jiang2021a, mei2017} which creates smaller structures. Many magnetohydrodynamic (MHD) and Particle-in-Cell (PIC) simulation works have investigated the two-dimensional or three-dimensional complex current sheet to discover the plasma condition and electromagnetic structures inside it \citep{wang2023, ye2023, daughton2011, nishida2013}. \added{And in highly-resolved resistive MHD studies, where a 2.5D current sheet is followed to transit to cascading reconnection, the role of interacting plasmoids at all resolvable scales was highlighted \citep{barta2011}.}

The understanding of particle acceleration and transport in the solar corona remains limited due to the difficulties of directly observing these particles as well as the challenges to link up the observations with the actual acceleration regions. Consequently, simulations have become the primary tool for investigating particle dynamics. Fully kinetic methods, such as PIC simulations, provide a self-consistent approach to study particle dynamics within reconnection fields \citep{dahlin2017}. However, the high computational cost of these methods restricts the simulation domain to limited scales, while a combination of PIC and MHD simulations remains a practical tool to model the standard scales of solar flares \citep{baumann2013}. To address larger-scale simulation situations, test particle methods are used in conjunction with MHD simulations, offering a computationally efficient alternative, albeit without feedback mechanisms. The motion of charged particles is fundamentally described by the Newton-Lorentz equations, and approximate analytical solutions have been derived for particle trajectories within modeled current sheet \citep{speiser1965}. \added{Additionally, the preferential acceleration of heavy protons, varying in both mass and charge, has also been investigated following the Lorentz evolution \citep{Kramolis2022}.} Given that the gyro-radius of electrons is approximately 1 meter, which is significantly smaller compared to the typical flare scale of $10^7$ meters, further simplifications are necessary to optimize computational resources. Particle transport equations are widely applied to model diverse physical processes from shock to reconnection \citep{guo2021a}. Recent investigations of flare loop-top sources have identified termination shocks as highly efficient particle acceleration sites by transport equations \citep{kong2022, kong2022a}. The Guiding Centre Approximation (GCA) with the adiabatic-invariant assumption \citep{northrop1963a} is frequently employed to investigate particle dynamics across electromagnetic fields \citep{karlicky2007, zhou2016, gordovskyy2019, gordovskyy2023, bacchini2024}. Nevertheless, feedback from particles has been added to the MHD equations to achieve more self-consistence \citep{drake2019, ruan2020, druett2024, seo2024a}. 

Notably, particle acceleration asymmetry has been investigated in a simplified three-dimensional current sheet configuration. When the guide field is sufficiently strong, electrons and protons are found to be accelerated in different directions, leading to distinct footpoint locations \citep{zharkova2004}. This asymmetric behavior has been demonstrated through full Lorentz equation calculations for particle motion, and later corroborated by two-and-a-half-dimensional PIC simulations \citep{karlicky2008}. However, the commonly observed conjugate footpoint sources in HXR observations are primarily attributed to accelerated electrons, which contradicts these theoretical findings where electrons alone would only produce a single-sided footpoint source. This discrepancy suggests the need for more realistic three-dimensional magnetic field configurations to properly reconstruct electron acceleration trajectories and their ultimate footpoint destinations.  Conjugate proton sources, mainly detected in gamma-ray observations \citep{hurford2006}, are beyond the scope of this study, which focuses on electron-dominated acceleration mechanisms.

In this study, we conduct a three-dimensional high-resolution MHD simulation to model the intricate structure of the current sheet. Based on the results of this MHD simulation, we perform test-particle simulations to explore the mechanisms of particle acceleration and transport originating from the fine-structured current sheet and ultimately reaching the lower solar atmosphere regions. In our simulations, electrons are injected into topologically complex, but non-evolving magnetic and electric fields. The primary aim of this research is to compare particle acceleration and transport in current sheet configurations with simple structures against those exhibiting tearing-mode instabilities. We place particular emphasis on analyzing the energy spectra and the dynamics of electrons that reach the lower boundary regions. By investigating these characteristics, we seek to reconstruct the ribbon sources observed on both sides of the current sheet in a three-dimensional situation.

The structure of this paper is organized as follows: Section \ref{sec:met} describes the methodology, including the MHD framework and test-particle approaches employed in our numerical simulations. Section \ref{sec:resmhd} presents a detailed analysis of current sheet evolution and its associated fine structures derived from MHD simulations. Section \ref{sec:restp} examines particle acceleration and transport within the current sheet, with particular emphasis on the influence of mini flux ropes on electron footpoint distributions. Finally, Section \ref{sec:discon} provides a comprehensive summary of our findings and discusses their implications.

\section{Methods} \label{sec:met}

\subsection{MHD Model}

In this study, we consider isothermal MHD equations:

\begin{eqnarray}
    & \partial_t \rho + \nabla \cdot (\rho \mathbi{v}) = 0,\\
    & \partial_t (\rho \mathbi{v}) + \nabla \cdot (\rho \mathbi{v} \mathbi{v} + p_{tot} \mathbi{I} - \mathbi{B} \mathbi{B}) = \rho \mathbi{g}, \\
    \label{eq:induction}
    & \partial_t \mathbi{B} = \nabla \cdot (\mathbi{v} \mathbi{B} - \mathbi{B} \mathbi{v} ), \\
    & p_{tot} = p + \frac{\mathbi{B} \cdot \mathbi{B}}{2},
\end{eqnarray}
where $\rho$ denotes the plasma density, $\mathbi{v}$ represents the plasma velocity, and $\mathbi{B}$ is the magnetic field. The total pressure is denoted by $p_{tot}$, while $p$ signifies the thermal pressure. These isothermal equations are complemented by the relation $p = c_{adiab} \rho$, where $c_{adiab}$ is a constant associated with the squared sound speed. To facilitate our simulations, we normalize all physical quantities to dimensionless units. The chosen scaling parameters are $10^9$ cm for length, $10^6$ K for temperature, and $10^9$ cm$^{-3}$ for number density, which are typical values for coronal conditions. Based on these, we can derive the dimensionless units for other physical quantities: density is 2.34 $\times 10^{-15}$ g~cm$^{-3}$, velocity is 116.45 km~s${^{-1}}$, pressure is 0.32 erg~cm$^{-3}$, magnetic field strength is 1.99 G, and the time is 85.87 s, noted as $t_0$. The gravitational acceleration decreases with height and is defined by $\mathbi{g} = -g_0 (R_{\odot}/(R_{\odot}+z))^2 \hat{\mathbi{z}}$, where $g_0 = 2.74 \times 10^4$~cm~s${^{-2}}$ is the acceleration on the solar surface, and $\hat{\mathbi{z}}$ is the unit vector in the z direction. The isothermal temperature is fixed at 1 MK, which determines the initial plasma density $\rho$ through the gravitational stratification. \added{Note that this isothermal assumption, along with the ideal version of the induction equation \ref{eq:induction}, implies that we leave relevant energetic processes (like Ohmic heating, or also thermal conduction, radiative losses) out and focus instead on achieving a high effective numerical resistivity that is capable of resolving current-sheet fine-structure \citep{mei2017}.}

The initial magnetic field configuration is constructed using the standard Titov and Démoulin modified model \citep{titov2014}. This magnetic structure is composed of two primary components. The first component is an MFR characterized by a major radius $R$, a minor radius $a$, toroidal flux $F$, and toroidal current $I$, with $I$ having a parabolic distribution along the minor radius. The second component consists of a pair of magnetic charges $q$, separated by a distance $L$ and located at a depth $d$ beneath the bottom boundary. For our simulation, the parameters are set as follows: $R$ = 2.4, $a$ = 0.3, $q$ = 7.5, $L$ = 1, and $d$ = 1. The values of $F$ and $I$ are determined based on external equilibrium conditions; however, to initiate the eruption of the MFR, we increase the toroidal current $I$ by a factor of 1.4. The adjustment is designed to facilitate the study of dynamic processes during the eruption phase. The magnetic flux rope is of negative helicity, as the directions of $F$ and $I$ are opposite.

These equations are solved using the Message Passing Interface Adaptive Mesh Refinement Versatile Advection Code \citep[MPI-AMRVAC;][]{xia2018a, keppens2023}. For the numerical scheme, we employ a three-step Runge-Kutta time discretization alongside the Harten-Lax-van Leer (HLL) Riemann solver, complemented by a fifth-order Weighted Essentially Non-Oscillatory (WENO) limiter. The computational domain is defined as a three-dimensional Cartesian box with sizes -4 $\leq$ $x$ $\leq$ 4, -4 $\leq$ $y$ $\leq$ 4, and 0 $\leq$ $z$ $\leq$ 8, so we look at a region of 80 Mm in size. The base resolution of the first AMR level is 48 $\times$ 48 $\times$ 48. We apply four AMR levels based on the default method, which emphasizes density gradients in MPI-AMRVAC. Additionally, two further AMR levels are applied using manual criteria to specifically refine the current sheet location: 

\[
\left\{
\begin{array}{l}
J \Delta / B > 0.25, \\
|\nabla(B^2/2)|\Delta / \rho > 5, \\
|(\mathbi{B} \cdot \nabla)\mathbi{B}| \Delta / \rho > 5,
\end{array}
\right.
\]
where $\mathit{J}$ is the current density and $\Delta$ is the highest resolution $\Delta x=\Delta y =\Delta z = \Delta \approx 52$ km. These criteria impose conditions less stringent than \cite{jiang2021a}. This approach allows our equivalent resolution to achieve 1536 $\times$ 1536 $\times$ 1536 within the current sheet region and \replaced{256 $\times$ 256 $\times$ 256}{384 $\times$ 384 $\times$ 384} in the surrounding regions, which is similar to the effective resolution setup in \cite{mei2017}. The boundary conditions incorporate a line-tied condition at the bottom boundary, where the velocity in the ghost cells is antisymmetric, and other quantities are extended using a fourth-order zero-gradient extrapolation. At the top boundary, a similar approach is applied, but the upward velocity is restricted to be positive. The side boundaries remain open, employing a constant extrapolation method for all quantities.

\subsection{Test Particle Model}

\begin{figure}
\includegraphics[width=1.0\textwidth]{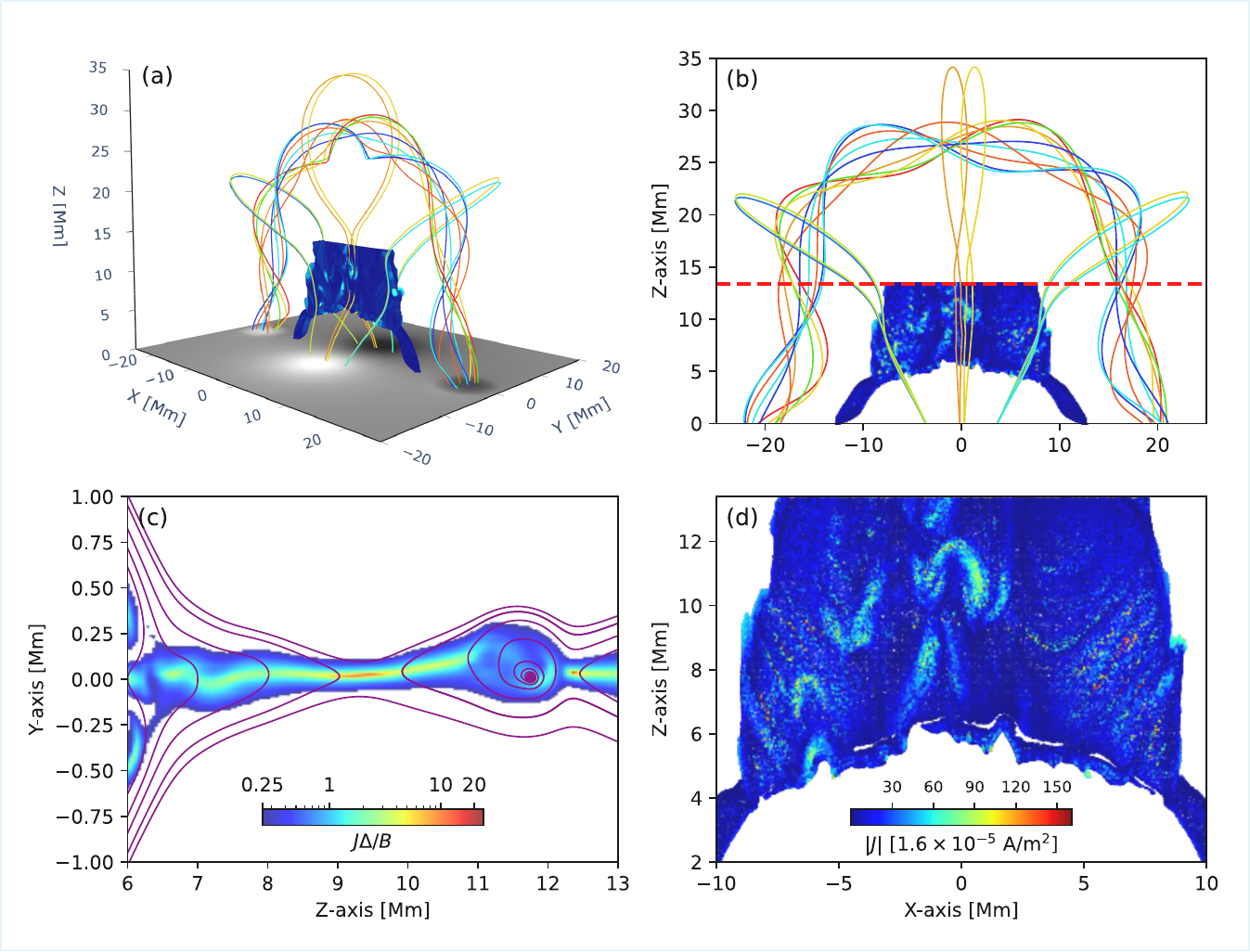}
\caption{The magnetic flux rope (MFR) and current sheet location defined using specific criteria at \(t = 4.4 \, t_0\). Panel (a) shows their relative position within the three-dimensional simulation box. Panel (b) projects them onto the XZ plane at \(y=0\).  \added{Panel (c) projects on to the YZ plane at \(x=0\) colored by the current sheet locator criteria, augmented with the reconnecting magnetic field lines.} The current sheet is colored by current density in panel \replaced{(c)}{(d)}.} 
\label{fig:mfrcs}
\end{figure}

To investigate the behavior of electrons originating from the current sheet, we utilize the test-particle module from MPI-AMRVAC. Given that the electron lifetime scale in the solar corona situation is significantly shorter than the evolution time scale of the MHD simulation \citep{bacchini2024}, we inject electrons into fixed magnetic and electric fields. The injected electrons follow a Maxwellian velocity distribution with a temperature $T$ = 1 MK, consistent with the isothermal temperature employed in our MHD simulations. Considering that the spatial and temporal scales of the MHD system are much larger than the electron gyro-radius and gyro-period in solar corona situation, we employ the GCA to evolve the electrons \citep{northrop1963a}. In the GCA framework, the particle motion along the magnetic field line is averaged over its gyration assuming an adiabatic motion. The parallel velocity along with the magnetic field line is given by $u_\parallel=\gamma \mathbi{v} \cdot \hat{\mathbi{b}}$, where $\gamma$ is the Lorentz factor, and $\hat{\mathbi{b}}$ is the unit vector along the magnetic field direction. The evolution equations focus on the guiding-centre $\mathbi{R}$, parallel velocity $u_\parallel$, and magnetic moment $\mu$, particularly in relativistic situations \citep{bacchini2024}:

\begin{equation}
\frac{d\mathbi{R}}{dt} = \frac{u_\parallel}{\gamma} \mathbi{b} + \mathbi{v}_E + \mathbi{v}_{\text{curv}} + \mathbi{v}_{\text{pol}} + \mathbi{v}_{\nabla B} + \mathbi{v}_{\text{rel}},
\end{equation}

\begin{equation}
\frac{du_\parallel}{dt} = \frac{q}{m} E_\parallel + a_{\text{curv}} + a_{\nabla B},
\end{equation}

\begin{equation}
\frac{d\mu}{dt} = 0,
\end{equation}
The parallel electric field is defined as $E_\parallel = \mathbi{E} \cdot \hat{\mathbi{b}}$. Addition to the parallel velocity $u_\parallel$, the motion of guiding-centre position is also evolved by the five drift velocity terms: $\mathbi{v}_E$, $\mathbi{v}_{\text{curv}}$, $\mathbi{v}_{\text{pol}}$, $\mathbi{v}_{\nabla B}$, and $\mathbi{v}_{\text{rel}}$ perpendicular to the magnetic field: 

\begin{equation}
\mathbi{v}_{\text{E}} = \mathbi{E} \times \mathbi{B} / B^2,
\end{equation}

\begin{equation}
\mathbi{v}_{\text{curv}} = \frac{mc\kappa^2}{qB} \mathbi{b} \times \left[ \frac{u_\parallel^2}{\gamma} (\mathbi{b} \cdot \nabla) \mathbi{b} + u_\parallel (\mathbi{v}_E \cdot \nabla) \mathbi{b} \right],
\end{equation}

\begin{equation}
\mathbi{v}_{\text{pol}} = \frac{mc\kappa^2}{qB} \mathbi{b} \times \left[ u_\parallel (\mathbi{b} \cdot \nabla) \mathbi{v}_E + \gamma (\mathbi{v}_E \cdot \nabla) \mathbi{v}_E \right],
\end{equation}

\begin{equation}
\mathbi{v}_{\nabla B} = \frac{\mu c \kappa^2}{\gamma q B} \mathbi{b} \times \nabla \left( \frac{B}{\kappa} \right),
\end{equation}

\begin{equation}
\mathbi{v}_{\text{rel}} = \frac{u_\parallel E_\parallel \kappa^2}{c \gamma B} \mathbi{b} \times \mathbi{v}_E,
\end{equation}
The coefficient $\kappa$ is derived from $\kappa = 1/\sqrt{1-v_E^2/c^2}$, and the conserved magnetic moment $\mu$ is defined by $\mu = mu_{\perp}^2/2B\kappa$. For the parallel direction, the acceleration items from the curvature and gradient of magnetic field are defined by:

\begin{equation}
a_{\text{curv}} = \mathbi{v}_E \cdot \left[ u_\parallel (\mathbi{b} \cdot \nabla) \mathbi{b} + \gamma (\mathbi{v}_E \cdot \nabla) \mathbi{b} \right], 
\end{equation}

\begin{equation}
a_{\nabla B} = -\frac{\mu}{m} \mathbi{b} \cdot \nabla \left( \frac{B}{\kappa} \right).
\end{equation}
The equations we consider in this study neglect all time derivative terms, relying on the fact that particle dynamics occur on much faster timescales than the MHD processes. The magnetic field in which particles are evolved is obtained from the MHD simulation through interpolation. The electric field is calculated using the relation $\mathbi{E} = -\mathbi{v} \times \mathbi{B} + \eta \mathbi{J}$, where $\mathbi{v}$ is the plasma velocity, $\mathbi{J} = \nabla \times \mathbi{B}$ is the current density, and $\eta$ is the resistivity. Note that our original isothermal MHD study did not involve resistivity, and as in \cite{mei2017} relied on numerical resistivity.  However, for the particle study we will adopt an anomalous resistivity that tries to model the beyond-MHD processes that the MHD model lack. The resistivity $\eta$ is considered anomalous depending on the current sheet location with the criteria where $\Delta$ is the local highest resolution, which is approximately 52 km:

\begin{equation}
\eta = \left\{
\begin{array}{ll}
10^{-5}, & \text{if } \frac{J \Delta}{B} > 0.25\\
0, & \text{otherwise}
\end{array}
\right.
\label{eq:cstp}
\end{equation}
In our simulation, the resistivity is set to $10^{-5}$ in the dimensionless unit, approximately 10 $\Omega \cdot $m in a specific location. The current density is on the order of $10^{-3}~\text{A} \cdot$m$^{-2}$, which results in an electric field of at least 0.01 $\text{V}\cdot$m$^{-1}$, exceeding the Dreicer field of about $0.0054~\text{V} \cdot$m$^{-1}$ for typical coronal parameters \citep{tsiklauri2006}. \added{And the manually applied resistivity is also inspired by the fact that earlier simulations indeed identified the need to go beyond a critical Lundquist number of about $10^4$ for chaotic island formation \citep{mei2017}.} Figure \ref{fig:mfrcs} illustrates the region of instantaneous, 3D current sheet meeting the criteria along with the surrounding magnetic field lines and the evolving MFR. The current sheet exhibits a quasi-planar structure, spatially resolved within the computational grid, with a thickness spanning approximately at most a dozen cells. From the surrounding magnetic field lines, which are undergoing reconnection, we can infer that the current sheet structure as judged from a cross-sectional 2D view resembles a two-dimensional reconnection scenario. The combination of all these reconnection sites forms the overall planar structure of the three-dimensional current sheet. We project this structure onto the XZ plane at $y = 0$ in Figure \ref{fig:mfrcs}(b). The red dashed line at $z = 1.4$ indicates that the region of interest is confined below the upward outflow region of reconnection, avoiding areas with unrealistically resistivity settings. In this projection, we recognize the standard spatial relationships among the MFR, the current sheet, and the surrounding reconnecting magnetic field lines.\added{Figure \ref{fig:mfrcs}(c) visualizes the current sheet from the side view along the $x$-axis, and we can find that the anomalous resistivity region is actually around the X-point and the long-stretched reconnection region. The magnetic field lines traced here are on the YZ plane at \(x=0\), where $B_z$ is approximately zero.} \replaced{Figure \ref{fig:mfrcs}(c)}{Furtherly, Figure \ref{fig:mfrcs}(d)} focuses on the details within the current sheet. Color-coded by $|\mathbi{J}|$ (the magnitude of the current density), we observe that the reconnection process is not uniform across the current sheet. Notably, there are elongated structures across different reconnection sites that exhibit stronger currents. These locations will be further discussed in the subsequent sections. 

\added{Moreover, the parallel energy change rate for GCA is presented here for further acceleration mechanism investigation \citep{zhou2016}:
\begin{equation}
\frac{1}{2}\frac{du_\parallel^2}{dt}=\frac{q}{m} E_\parallel u_\parallel+a_{\text{curv}} u_\parallel+a_{\nabla B} u_\parallel.
\label{eq:acc}
\end{equation}
Under the adiabatic invariant assumption, the change in the perpendicular energy is determined solely by the magnetic field strengths at its initial and final positions. Consequently, our analysis is centered on the parallel energy rate, which is predominantly attributable to two mechanisms: acceleration by the parallel electric field and curvature accelerations. The third term represents the conversion of parallel energy into perpendicular energy while conserving the total energy, a phenomenon that can be interpreted as a manifestation of the mirror effect. Considering that electrons originate in the current sheet and terminate at the bottom boundary regions where the magnetic field strength is unambiguously increasing, the net outcome of this third term is a deceleration of the particle parallel velocity.}

\section{Results of MHD Simulation} \label{sec:resmhd}

\subsection{Evolutions of MHD Simulation}

\begin{figure}
\includegraphics[width=0.7\textwidth]{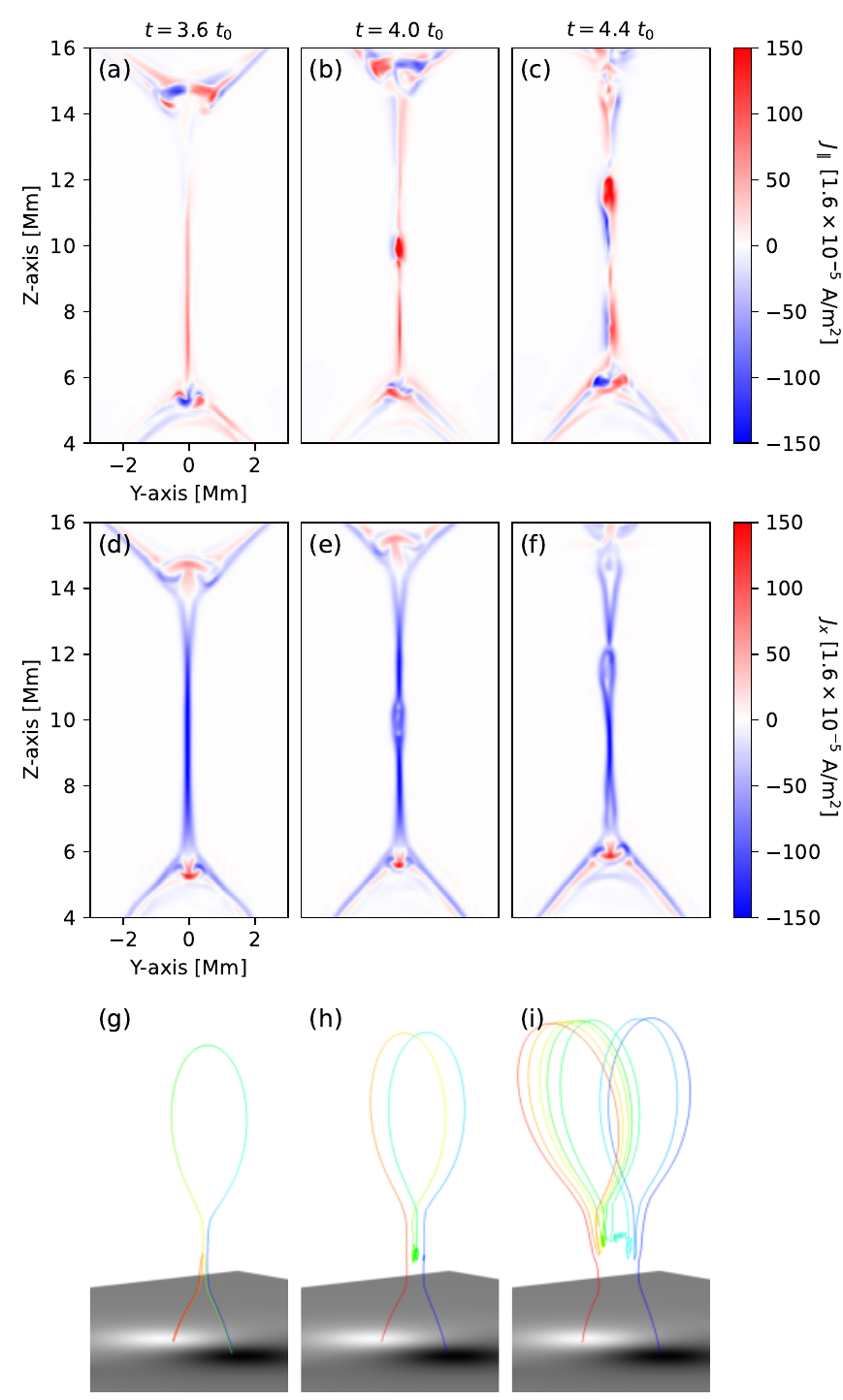}
\caption{Snapshots showing the evolution of (a--c) the parallel current \added{density}, (d--f) current density \added{along the $x$-axis}, and (g--i) magnetic field lines at $t=$3.6 $t_0$, 4.0 $t_0$, 4.4 $t_0$, respectively. The magnetic field lines are integrated around the 2D slice position.}
 \label{fig:mhdevol}
\end{figure}

At the onset of the MFR's rise, mini flux ropes do not form. This is because the tearing instability is expected to develop only when the current sheet's aspect ratio exceeds a critical threshold, a condition achieved in the simulation under high Lundquist number, with resistivity solely governed by numerical effects. The current sheet comprises multiple two-dimensional-like reconnection sites on the plane $y=0$ approximately. We illustrate in Figure \ref{fig:mhdevol} one two-dimensional-like reconnection scenario in the middle of the MFR from an YZ projection at $x=0$, which shows a cross-sectional view perpendicular to the current sheet structure itself, akin to the scenario of the CSHKP model. The evolution of the reconnection process is shown by the parallel current density, $J_{\parallel}=\mathbi{J} \cdot \hat{\mathbi{b}}$, as shown from $t$=3.6~$t_0$ to 4.4~$t_0$ in the upper row of Figure \ref{fig:mhdevol}. The sign of the current helicity, $\mathbi{J} \cdot \mathbi{B}$, remains consistent with the parallel current, deviating only from a constant $|\mathbi{B}|$ in the aspect of magnitude.

In Figure \ref{fig:mhdevol}(a), it is observed that magnetic islands do not appear in the two-dimensional-like reconnection scenario when $t = 3.6\, t_0$. The direction of $J_\parallel$ in the current sheet slice aligns with the magnetic field lines ($J_{\parallel}>0$), yet it contrasts with the negative helicity of the erupting MFR. In Figure \ref{fig:mhdevol}(b), the emergence of magnetic islands is evident, with a negative parallel current region developing in the central top region of the current sheet. By $t = 4.4\, t_0$, the current sheet has bifurcated into regions of positive and negative parallel current in Figure \ref{fig:mhdevol}(c). The emergence of the magnetic islands, along with the simultaneous existence of both antiparallel and parallel current, provides strong evidence that the three-dimensional current sheet significantly deviates from the two-dimensional CSHKP model, even when analyzed through a two-dimensional slice. The antiparallel current structures emerge simultaneously with the formation of magnetic islands, indicating a strong temporal correlation between these two structures, which can be interpreted as a consequence of tearing instability, even though our 3D isothermal MHD simulation only has numerical resistivity at play. 

Figures \ref{fig:mhdevol}(d)--(f) depict the evolution of $J_x$, the current in the $x$-axis direction in the central MFR region, the same region as Figures \ref{fig:mhdevol}(a)--(c). It is expected that $J_x$, which mimics the reconnection generated current, maintains the same sign as the toroidal current of the MFR during the emergence of magnetic islands. This aligns with expectations based on the uniform chirality direction of the field's curl in the basic configuration of reconnecting magnetic field lines. It suggests that small-scale field changes, associated with tearing instability, occur rather than large-scale changes, which would result in the sign inversion of the parallel current.

Furthermore, Figures \ref{fig:mhdevol}(g)--(i) illustrate the magnetic field lines around the current sheet. In Figure \ref{fig:mhdevol}(g), prior to the formation of magnetic islands, the magnetic field lines lie approximately in the plane of $x=0$, indicating that the magnetic configuration is just like the standard flare scenario of a two-dimensional CSHKP model. However, in Figure \ref{fig:mhdevol}(h), the emergence of mini flux ropes is evidenced by the small rope-like structures traversing different reconnection sites on the current sheet. The magnetic islands are projections of the mini flux ropes in a two-dimensional YZ plane in Figures \ref{fig:mhdevol}(a)--(f). In Figure \ref{fig:mhdevol}(i), the mini flux rope extends with its axis oriented mainly in the near-planar current sheet, highlighting the full 3D nature of reconnecting magnetic fields. Thus, the dominant reconnetion process is a three-dimensional configuration, totally different from the two-dimensional CSHKP model. It is important to note that the tearing instability is observed to have occurred at the snapshot corresponding to \(t=3.6\ t_0\). However, it does not correspond to the location depicted in the cross-sectional view. Due to the interval between \added{saved} snapshots, we did not conduct a detailed analysis to determine the precise time of onset across the entire domain. The earliest snapshot in which the tearing instability can be traced is at \(t=\ 2.0\ t_0\). Here we select \(t=4.4\ t_0\) for further detailed analysis, as prior timesteps lack sufficient numerical resolution to adequately resolve fine structures, which are distinctly captured at this stage.

\subsection{Mini Flux Ropes in the Current Sheet}

\begin{figure}
    \includegraphics[width=1.0\textwidth]{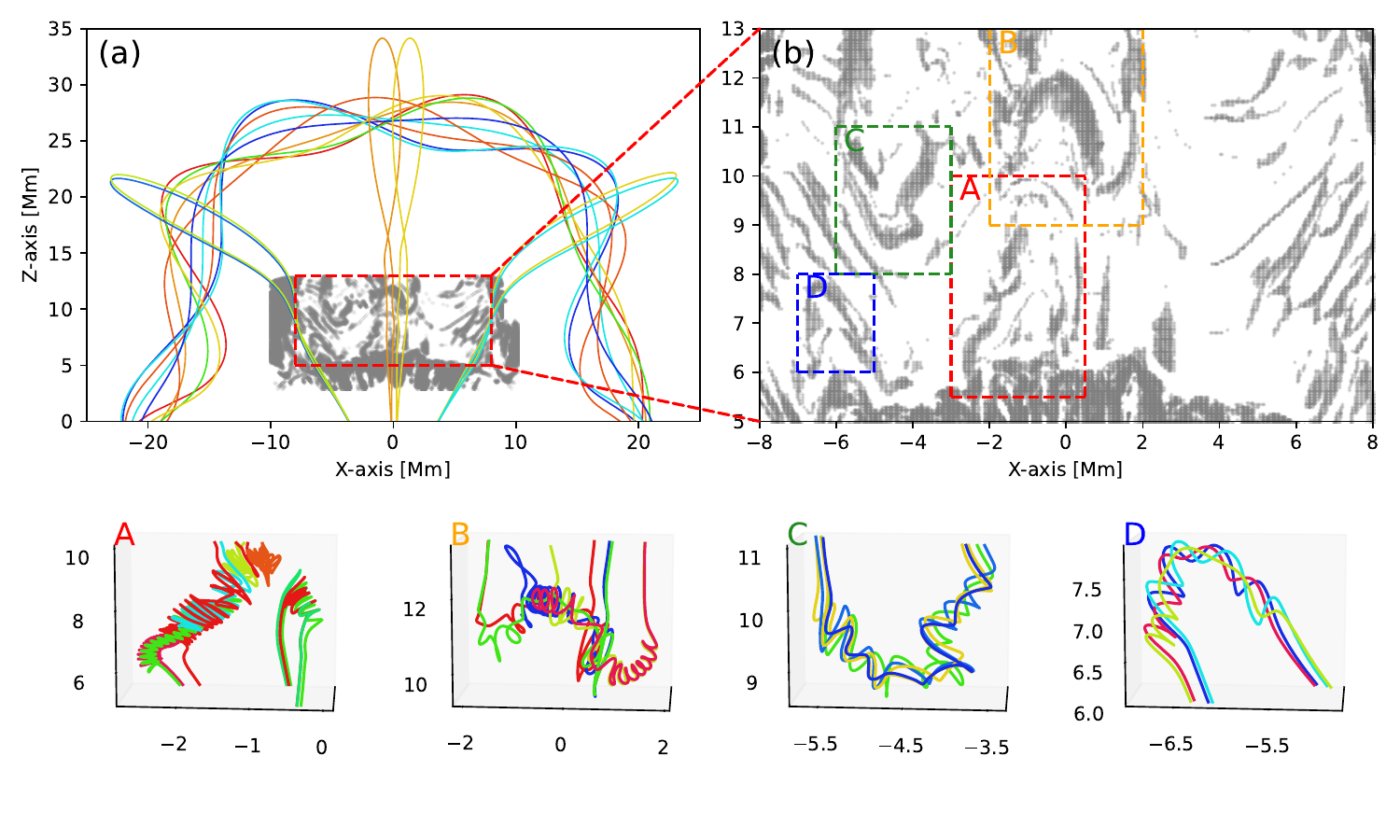}
    \caption{Distribution of O-type \replaced{points}{field line sites} within the current sheet (a,b) and the mini flux ropes around these reconnection points (A--D). Panel (b) enlarges the red box region in panel (a) to show the detailed distribution.}
    \label{fig:mmfrs}
\end{figure}

Mini flux ropes are not confined to a single subregion of the current sheet. At \(t = 4.4\, t_0\), we use the LoRD (Locate Reconnection Distribution) method to identify mini flux ropes within the selected current sheet region \citep{wang2023}, assuming all reconnection sites lie near the plane-like current sheet area. The method utilizes a modified Jacobian matrix to identify different types of reconnection. The results indicate O-type \replaced{reconnection sites}{field line sites}, confirming the presence of mini flux ropes, as depicted in Figure \ref{fig:mmfrs}(a). We then enlarge the central region of interest to exclude any positions related to reconnection outflow regions. The region outlined by the red dashed box is presented in Figure \ref{fig:mmfrs}(b), where we observe elongated stripe structures, resembling the high current regions in Figure \ref{fig:mfrcs}. The other points are falsely flagging outflow regions of reconnection, where the magnetic field configuration mimics O-type \replaced{reconnection}{field line sites} in the LoRD method. We identify and label four regions of the correctly identified \deleted{O-type} points as $\mathit{A}$, $\mathit{B}$, $\mathit{C}$, and $\mathit{D}$. By integrating magnetic field lines around them, we identify four mini flux ropes in the bottom panels of Figure \ref{fig:mmfrs}. From the connectivity of these mini flux ropes, we infer that they form near the upward or downward outflows of reconnection, similar to what has been found in \citet{ye2023}. Among the mini flux ropes, \(\mathit{B}\) and \(\mathit{C}\) connect to the upper magnetic field configuration, while \(\mathit{D}\) connects to the lower configuration. However, the linkage of mini flux rope \(\mathit{A}\) is more complex, as some magnetic field lines link downward, while others connect upward to mini flux rope \(\mathit{B}\).

\section{Results of Test Particle Simulation} \label{sec:restp}

We now set forth to explore the consequences of having detailed mini flux rope structures in the current sheet on test particle acceleration and transport. We inject about 5 million electrons into the current sheet region, depending on the criteria of Equation (\ref{eq:cstp}). These electrons follow a Maxwellian velocity distribution at 1 MK and are traced until $0.1 \ t_0$ ($\sim$8 s), but we keep the electromagnetic fields unchanged from the selected MHD snapshot at \(t = 4.4\, t_0\). To conserve computational resources, the domain for test particles is a subregion of the MHD domain, with sizes -1.34 $\leq$ $x$ $\leq$ 1.34, -1.34 $\leq$ $y$ $\leq$ 1.34, and 0 $\leq$ $z$ $\leq$ 1.34. Technically, we kept the instantaneous AMR grid block hierarchy in this subregion intact, but reordered the space-filling curve through all the blocks such that we can perform the test particle study in a separate restart with MPI-AMRVAC that handles this subregion only. The domain is sufficiently large to study especially the acceleration and transport of electrons injected in the current sheet. The electrons are traced either until the time limit \(0.1 t_0\) (\(\sim 8\ s\)) is reached or their trajectory is collected up to the point when they encounter the boundary of the simulation box. The electrons precipitating on the bottom boundary, corresponding to the HXR footpoint sources, are of particular interest. We do not analyze electrons reaching the upper boundary here because the magnetic field configuration in the eruping flux rope structure is entirely closed, and this aspect of loop-trapped particle acceleration has been studied before by \cite{pinto2016}. We therefore reduced the upper height of the simulation box to 1.34 to prevent electrons from crossing into the MFR's loop. We note that for all electrons inserted in the current sheet region, no electrons escape from the side boundaries, so all end up either fully traced during the timeframe studied, or they end up at top or bottom boundaries.

Prior to presenting our particle simulation results, we recall the asymmetric acceleration and transport trajectories of electrons and protons within a simplified current sheet configuration, which directs them to different footpoints at the lower boundary region when the ratio \(B_y/B_z\) is relatively high \citep{zharkova2004}. This \(B_y/B_z\) criterion can be interpreted as qualitatively analogous to the guide field in the 3D reconnection sites under the GCA framework, which is theoretically sufficient to maintain the magnetization electrons. In this context, thermal particles, predominantly driven by direct current within the current sheet, undergo acceleration in the direction along or opposite to the parallel current along the magnetic field lines. This acceleration mechanism demonstrates that particle trajectories are primarily governed by the direction of the parallel current and the particles' electric charge, resulting in an asymmetric footpoint distribution between electrons and protons. As illustrated in Figure \ref{fig:mhdevol}(a), before the occurrence of tearing instability, the parallel current maintains a uniform direction along the magnetic field in the current sheet's central region. Under these conditions, particles of identical charge are confined to unidirectional motion, either parallel or antiparallel to the magnetic field, resulting in a single-sided source at the boundary. This phenomenon occurs because the drift velocity effects are negligible compared to the dominant parallel electric field. However, this theoretical analysis is confined to a simplified reconnecting current sheet configuration. As shown in top panels of Figure \ref{fig:mhdevol}, we witness the emergence of antiparallel currents following the development of tearing instability, and this proves to be crucial in introducing complexity to particle motion.

\subsection{Electrons in the Mini Flux Ropes} 

\begin{figure}
    \includegraphics[width=.9\textwidth]{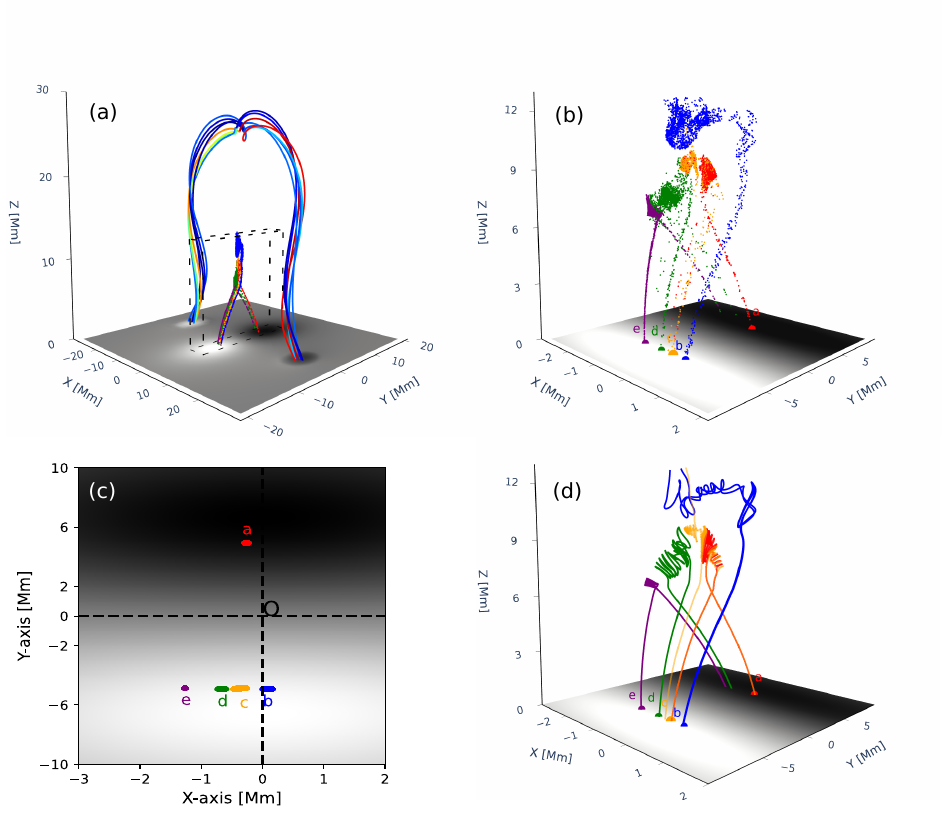}
    \caption{Trajectory of high-energy electrons ($\geq$ 25 keV) precipitating on the bottom boundary (a, b), their precipitation positions (c), and magnetic field lines around the trajectory (d). Panel (b) enlarges the region within the dashed black box in panel (a) for clarity.}
    \label{fig:bottom}
\end{figure}

We analyze the electrons that ultimately reach the bottom boundary, focusing on the high-energy ones from all the collected electrons. The energy threshold for selection is set above 25 keV, a typical nonthermal energy range for HXR footpoint sources in observation. Out of the 5 million electrons injected in total, approximately 1 million are collected at the lower boundary, with roughly 20,000 of these electrons exceeding the 25 keV energy threshold. We identify five discrete sources in Figure \ref{fig:bottom}(c), labeled a, b, c, d, and e, which represent all the bottom precipitation electron sources exceeding 25 keV. Source a is located on the positive $y$-axis, while the other four are on the negative $y$-axis. The presence of sources on both sides already indicates that it is modified from the preferential one-sided electron acceleration scenario in the simplified current sheet.

To understand electron behavior, we first present an overview of the trajectories which we can trace back from these sources within the entire simulation box, including MFR, to highlight their relative spatial locations, as shown in Figure \ref{fig:bottom}(a). We find that the trajectories that lead to energetic impact at the bottom are located in the central part of the current sheet. We then enlarge the view to focus on the trajectories for a clearer analysis in Figure \ref{fig:bottom}(b). We find that the traceback trajectory for each source is quite distinct, indicating differences in their acceleration sites. Each site corresponds to a single source on one side, suggesting that the tendency to accelerate electrons in one direction toward one side is still retained. Interestingly, some trajectories extend partially to the opposite side where no corresponding source exists. This electron bounce is related to the magnetic mirror effect, indicating inefficiency in accelerating electrons to the opposite side within the specific magnetic field structure. Additionally, we find trajectories similar to the mini flux ropes in terms of shape and location. In Figure \ref{fig:bottom}(d), by integrating some typical magnetic field lines from points around these trajectories, we discover that the magnetic field lines align with the mini flux rope regions in Figures \ref{fig:mmfrs}$\mathit{A}$ and $\mathit{B}$. This demonstrates the efficiency of mini flux ropes in accelerating electrons, as only those electrons accelerated from mini flux ropes exceed 25 keV. The two mini flux ropes situated in the central part of the current sheet, where the strongest reconnection and intense current density occur at \(t=\ 4.4 t_0\), are capable of accelerating electrons to such high energies. The five sources correspond to subsets of the two mini flux ropes, and each source correlates with the magnetic field line connection at the bottom boundary, implying that electron trajectories align with the magnetic field lines of the mini flux ropes.

\begin{figure}
    \includegraphics[width=.9\textwidth]{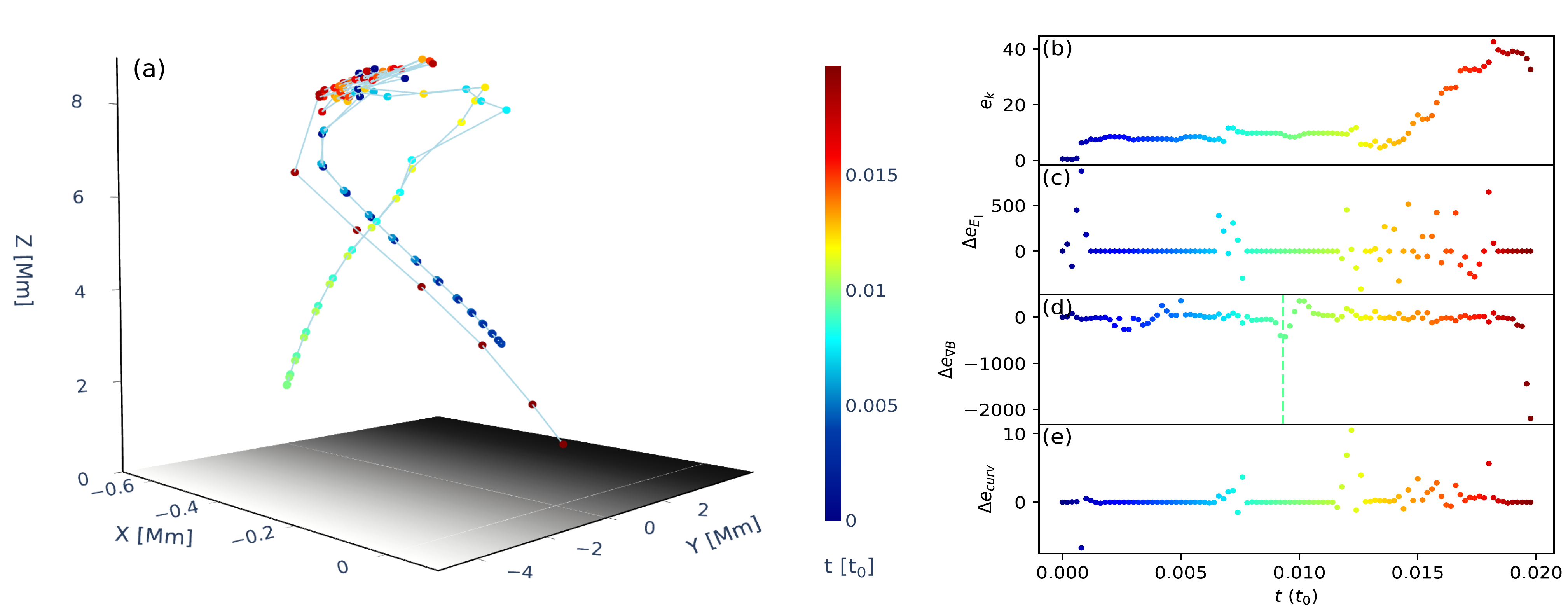}
    \caption{(a) Trajectory of a single electron that eventually reaches the bottom boundary source position a and (b) its kinetic energy and \added{parallel} energy change during the trajectory. Both trajectory and energy are colored according to time. Note that the acceleration from parallel electric field only happens when it is inside the current sheet structure.}
    \label{fig:single}
\end{figure}

To further support this interpretation of all energetic sources as originating in mini flux ropes, we analyze a single electron from source a, as displayed in Figure \ref{fig:single}. In Figure \ref{fig:single}(a), the electron's trajectory is depicted using points at different times, with each point colored to illustrate the evolution of the electron's path over time. Initially, the particle moves to the opposite side from its final source location but is bounced back to the mini flux rope due to the strong magnetic mirror effect. This effect is illustrated in Figure \ref{fig:single}(d) at the time marked by the green dashed line, indicating the strong deceleration from the significant gradient of magnetic field. For nearly half of the lifetime between the injection and the precipitation, the electron resides within the mini flux rope, undergoing cycles of acceleration and deceleration in Figure \ref{fig:single}(c). This behavior results from a mix of positive and negative parallel current values in Figure \ref{fig:mhdevol}(c). However, the overall effect of the parallel current across the mini flux rope is to accelerate electrons in one direction, ultimately resulting in a high-energy electron precipitating at one side of the bottom boundary, evident in the kinetic energy enhancement shown in Figure \ref{fig:single}(b). Additionally, in Figure \ref{fig:single}(e), curvature acceleration is not effective due to the absence of a structure large enough to trap electrons for an extended period, as the mini flux ropes are relatively short. Ultimately, we find that the dominant mechanism for accelerating the single electron is the parallel electric field as expected. Note that we took the anomalous resistivity prescription such that no further parallel acceleration is active when the electron leaves the current sheet.

\begin{figure}
    \includegraphics[width=.9\textwidth]{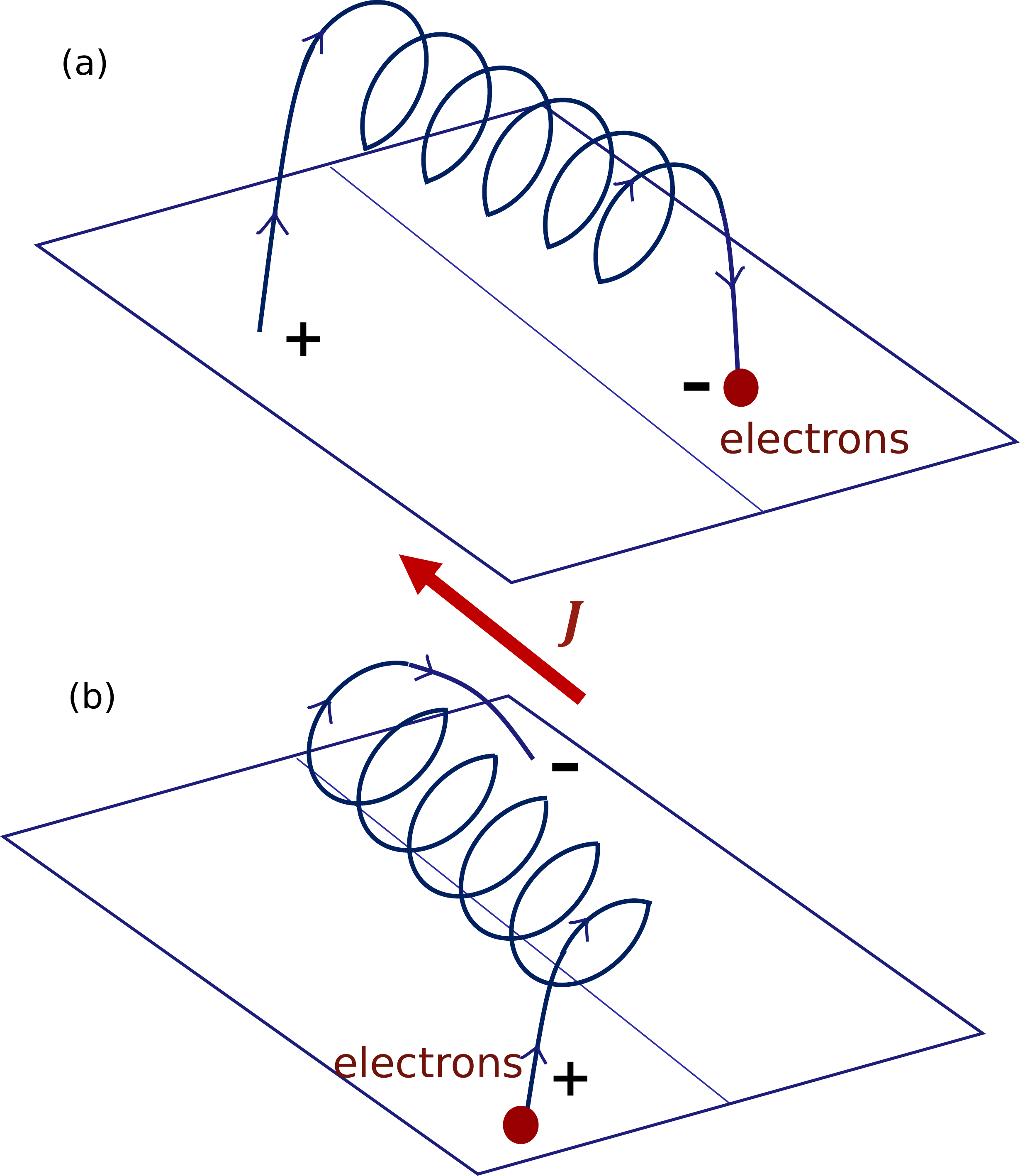}
    \caption{Schematic illustration of mini flux ropes with different chiralities within a fixed current and background magnetic field framework. The upper panel shows a dextral mini flux rope with left-handed twisted fined lines and negative magnetic helicity, while the bottom panel shows a sinistral one with right-handed twisted fined lines and positive magnetic helicity.}
    \label{fig:cartoon}
\end{figure}

For the five individual sources corresponding to subsets of mini flux ropes, the acceleration of electrons in a specific direction can be explained by the magnetic field helicity of the mini flux ropes. Figures \ref{fig:mhdevol}(a)--(c) show that the current sheet has a positive magnetic helicity since $J_{\parallel}$ is positive initially, and mini flux ropes with both positive and negative helicity are formed in a later time. Figures \ref{fig:mhdevol}(d)--(f) further show that the sign of $J_x$ remains unchanged. The electrons are accelerated in the reverse direction to the current due to its negative charge. With opposite magnetic helicity, mini flux ropes can exhibit two types of axis chirality as shown in Figure \ref{fig:cartoon}. For the dextral one, the flux rope axis magnetic field points to the right when an observer stands on the positive polarity, the flux rope field lines are left-handed twisted, and the magnetic helicity is negative as shown in Figure \ref{fig:cartoon}(a). Conversely, in Figure \ref{fig:cartoon}(b), the mini flux rope axis field points to the left when an observer stands on the positive polarity, the flux rope field lines are right-handed twisted, and the helicity is positive. This connectivity directly affects the side of the final precipitation. Electrons are consistently accelerated along the reversed current direction towards the negative polarity in negative helicity and towards the positive polarity in positive helicity, leading to two potential precipitation locations. This scenario demonstrates a unidirectional tendency to drive electrons within each mini flux rope based on their helicity. The coexistence of mini flux ropes with opposite helicity supports the presence of double ribbon sources at the bottom boundary. In Figure \ref{fig:bottom}(d), we see that the five sources fit this scenario: source a aligns with the first case in Figure \ref{fig:cartoon}, whereas sources b, c, d, and e align with the second case. Notably, the main body of mini flux rope a wraps around the thinner axis of c, exhibiting a negative helicity configuration opposite of c. The opposite dextral mini flux rope is the result of tearing instability, which facilitates particle transport toward the conjugate footpoint. Notably, the observed conjugate double-source configuration does not arise from a single two-dimensional reconnection process, but instead manifests as a result of particle acceleration within mini flux ropes possessing opposite helicity.

\subsection{Energy Spectrum}

\begin{figure}
    \includegraphics[width=.9\textwidth]{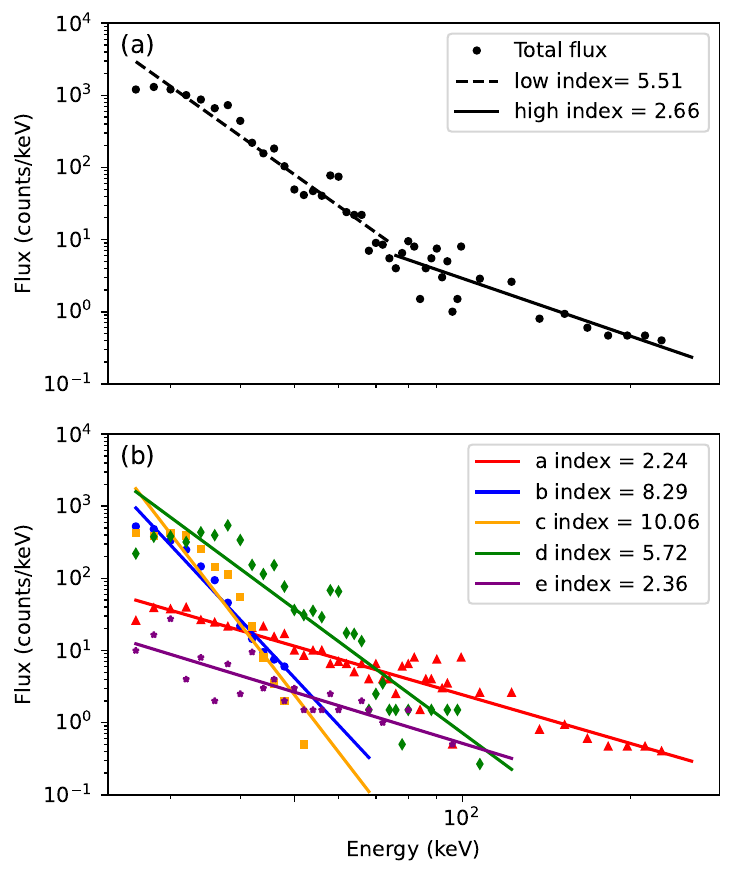}
    \caption{Spectrum of electrons precipitated on the bottom boundary at \(t = 4.4 \, t_0\). (a) Spectrum of the total electron population. (b) Spectra corresponding to those that end up as the five electron sources, colored to match Figure \ref{fig:bottom}.}
    \label{fig:spectrum}
\end{figure}

Figure \ref{fig:spectrum} shows the bottom precipitated electron energy distribution $f(E)=dN/dE$. We observe that the energy distribution of all electrons reaching the bottom exhibits a broken power law spectrum in Figure \ref{fig:spectrum}(a). The index for lower energies is 5.51, while it is 2.66 for higher energies. In Figure \ref{fig:spectrum}(b), we categorize all electrons into five groups depending on their trajectories as described above. We find that the previous source regions labeled as a and d contribute the most to the dominant energy, as their counts are significantly larger than the others, forming the higher energy and lower energy parts of the broken power law, respectively. Notably, electrons accelerated by mini flux ropes with negative helicity generated from the tearing instability achieve higher energies and exhibit harder spectral indices. In contrast, other acceleration regions, dominated by original positive helicity flux ropes, primarily contribute to the lower-energy particle population. This distinction provides a physical explanation for the break in the broken power law spectrum, directly linking it to the magnetic reconnection process. Thus, it shows that electrons are accelerated and transported to two main sources, located on different sides. This forms a double source structure comparable to observational results. The varying efficiency in accelerating electrons leads to different power law indices in different sources, with the two main mini flux ropes contributing to the broken power law spectrum observed in the energy distribution of all electrons. The asymmetry in conjugate footpoints, evident in both flux and spectral indices, aligns with previous observations from \citep{Jin2007}.

\subsection{Acceleration Mechanisms}

\begin{figure}
    \includegraphics[width=.9\textwidth]{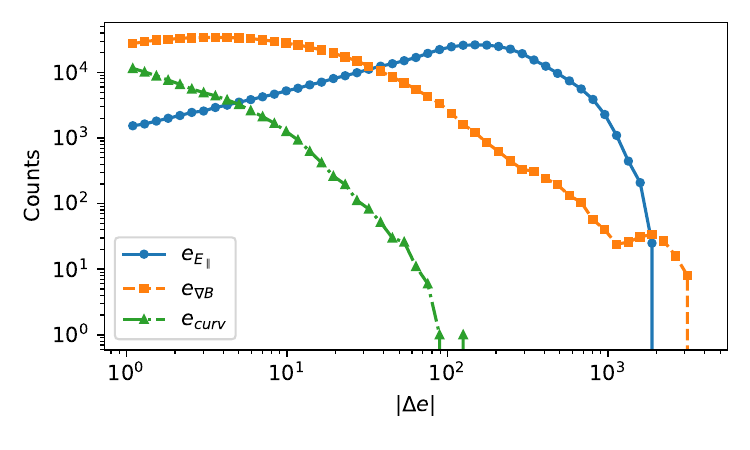}
    \caption{Histogram of \added{parallel} energy changes from all trajectory points in Figure \ref{fig:bottom}. Blue solid line and circles represent the parallel electric field term, orange dashed line and squares indicate the magnetic field gradient term, and green dash-dotted line and triangles show the curvature acceleration term.}
    \label{fig:mechanisms}
\end{figure}

The acceleration mechanisms in the simulation are investigated by analyzing them through a histogram. In Figure \ref{fig:mechanisms}, we calculate the contribution of various acceleration factors from all trajectory points, \added{based on the three terms in Equation (\ref{eq:acc}}), focusing on electrons that achieve final energies exceeding 25 keV. As expected, the parallel electric field emerges as the dominant acceleration mechanism, which is expected as the main acceleration sites are located in the current sheet with non-zero resistivity.  The magnetic mirror effect, however, cannot be ignored, as its maximum contribution is of the same magnitude as that of the parallel electric field. However, the net effect of the magnetic gradient is always deceleration in the acceleration driven by reconnection because the magnetic field in the reconnection sites is always greatly lower than that at the bottom. The curvature term contributes significantly less, which aligns with our single particle analysis in Figure \ref{fig:single}(d). The restricted contribution from the curvature term may arise from the limited length of mini flux ropes or their relatively strong internal guide fields \citep{dahlin2017}.

\section{Conclusions and Discussions} \label{sec:discon}

In this study, we explore particle acceleration within a three-dimensional large-scale current sheet under an erupting MFR, focusing particularly on the fine structures inside the current sheet, known as mini flux ropes. Particle motion is modeled using the GCA with electric fields described by \(\mathbi{E}=-\mathbi{v} \times \mathbi{B}+ \eta \mathbi{J}\), where anomalous resistivity is determined by specific criteria identifying the current sheet. Both the MHD simulation and the test particle simulation are conducted using MPI-AMRVAC where the latter is performed on a subregion of the full MHD setup, focusing on the current sheet and how it can accelerate particles downwards.

We begin by analyzing the evolution of the central slice of the current sheet. It is observed that mini flux ropes form only as expected in high Lundquist number settings, and as soon as its aspect ratio exceeds some threshold. Their emergence disrupts the uniform direction of parallel current along the magnetic field, as well as the magnetic helicity, although the overall central reconnection-generated current direction along the $x$-axis remains unchanged. This disruption is attributed to small-scale magnetic field changes, leaving the overall current direction unchanged. Additionally, mini flux ropes alter the simplified two-dimensional reconnection magnetic configuration, linking multiple reconnection sites.

Utilizing a snapshot of the MHD results, we inject electrons into the simulation box with velocities following a Maxwellian distribution at a temperature of 1 MK. Assuming the particle lifetime is much shorter than the MHD background's variation timescale, we find that mini flux ropes efficiently accelerate electrons with high local current densities that provide intense electric fields. The final positions and trajectories of electrons on the bottom boundary reveal that each precipitation corresponds to a mini flux rope. The magnetic helicity of these flux ropes determines the precipitation side, with negative and positive helicity mini flux ropes associated with negative and positive polarity precipitations, respectively. Notably, four footpoint sources on the positive polarity correspond to mini flux ropes with positive helicity, persisting from the tearing instability phase, and the mini flux ropes with negative helicity form after the tearing instability occurs. This magnetic configuration with both positive and negative helicity enables the electric field-driven electrons to be directed toward footpoints on both sides. In this scenario, the conjugate footpoint sources do not originate from the same field line, they instead arise from paired mini flux ropes with opposite helicity. The spatially fragmented acceleration of electrons aligns with the discovery of multiple acceleration sources for both HXR and radio emissions reported in \cite{bhunia2025}. 

The HXR observations routinely reveal a broken power-law distribution in the proton spectrum, implying a similar spectral break in accelerated electrons \citep{holman2003, alaoui2019}. In our simulation, the energy distribution of bottom-precipitated electrons exhibits a broken power law with a softer index at lower energies and a harder index at higher energies, which exhibits a spectral hardening \citep{kong2013, ning2019}. Detailed analysis of electrons from five ejection positions indicates that this distribution primarily arises from two subsets of mini flux ropes with different magnetic helicity, leading to double sources on both sides. The break energy in the power law spectrum originates from distinct reconnection processes within mini flux ropes, which are intrinsically linked to the evolution of tearing instability. Consequently, our findings suggest the existence of two strong electron sources, comparable to the standard flare model, alongside four weaker sources. High-energy electron acceleration is dominated by parallel electric fields, while curvature acceleration plays a minor role due to the limited length of mini flux ropes, which cannot efficiently trap the electrons.

Our work highlights the impact of three-dimensional magnetic field configurations on electron dynamics within a large-scale current sheet. We identify characteristic electron behaviors and energization patterns influenced by magnetic structure variations. However, several challenges remain unresolved. The most critical question concerns the occurrence of mini flux ropes with different magnetic helicity. Most mini flux ropes retain original positive magnetic helicity, leading to the formation of four electron sources on the bottom boundary. In contrast, the emergence of mini flux ropes with negative helicity is observed following the tearing instability. Further investigation is needed to understand the mechanisms governing the production of opposite magnetic helicity within the current sheet. \added{The magnetic configuration in our simulation is achieved by reversing the guide field while maintaining the direction of the current. In contrast, the reversal of the current direction can also be observed in 2.5D simulations when plasmoids merge and interact, forming secondary plasmois which represent finer fractal structures inside the mini flux ropes \citep{barta2011, karlicky2011}. Due to the limitations in resolution and the challenges of precisely controlling resistivity in our 3D simulation, we were unable to resolve and demonstrate substructure when mini-flux ropes interact to form further fractal structures. However, their potential presence could result in more complex and fragmented sites of particle acceleration, as well as a more intricate distribution of deposition sites of accelerated electrons in the flare footpoint regions.} Furthermore, the original positive helicity of the current sheet differs from the negative helicity of the MFR. This discrepancy could be attributed to the rotation of the MFR.

In our simulation, we focus on electrons accelerated within the planar current sheet located beneath the erupting MFR, with particular attention to the downward-propagating energetic electrons. While reconnection between the flux rope's magnetic field lines and the surrounding ambient field may also serve as a potential acceleration site, the electrons accelerated in such regions, along with upward-propagating electrons from the current sheet, are expected to orbit the flux rope and ultimately reach its footpoints, as described in \cite{pinto2016}. However, a detailed investigation of these processes lies beyond the scope of the present study, as the dominant reconnection in our model is localized to the planar current sheet beneath the erupting MFR. This contrasts with the configuration proposed by \cite{pinto2016}, wherein reconnection sites are concentrated around the flux rope's toroidal field lines — a configuration hypothesized to primarily accelerate electrons and transport them to the MFR footpoints.

\added{The acceleration mechanisms in our test particle simulation are not fully applicable to the kinetic scale, primarily due to the limitations in resolution, which lead to larger fine structures and unresolved secondary plasmoids. The typical dissipation scale can be estimated by the ion inertial length $d_i\equiv \frac{c}{\omega_{pi}}$, which is approximately 10 m in solar corona conditions \citep{buchner2006,Kramolis2022}. In our 3D MHD simulation, the finest achieved resolution is 52 km, which is at least three orders of magnitude larger than the typical dissipation scale. Resolving physics up to kinetic scales would require a $10^6(1000^2)$ reduction in cross-sectional area, and the scale along the guide field would also be similarly reduced. This difference implies that the acceleration and transport processes at the kinetic scale would deviate from those observed in our simulation. Specifically, particles would be accelerated by the parallel electric field within smaller spatial regions and would spend less time in a single acceleration site. However, finer-scale structures are also expected to generate more intense current densities and possibly enable secondary acceleration mechanisms due to the increased complexity and fragmentation of acceleration sites. These effects could further enhance the contribution of parallel electric field acceleration. Additionally, other acceleration mechanisms are likely to benefit from the finer structures present at the kinetic scale. For instance, in regions where the cascading of secondary plasmoids is fully developed, particles are expected to undergo more efficient Fermi-type acceleration \citep{fu2006a,li2015,li2019c,Kramolis2022}. One potential approach to bridge the gap between fully developed kinetic-scale cascading and MHD simulations is the use of  a downscaling factor to the spatial coordinates, resulting in a shifted kinetic scale by assuming self-similarity in the cascading process \citep{Kramolis2022}. However, this approach comes with the drawback of losing the ability to capture particle trajectories within a broader and more realistic physical region.} 

The energy threshold of 25 keV for electrons was adopted in our simulations for two purposes: (1) to align with the typical low-energy cutoff of nonthermal electrons, and (2) to isolate populations accelerated exclusively within mini flux ropes. While some electrons are energized via direct acceleration in X-line reconnection structures without subsequent trapping in flux ropes, none of these untrapped populations exceed 25 keV. These unidirectionally accelerated electrons, directed toward the positive polarity and consistent with our pre-simulation theoretical framework, exhibit significantly lower acceleration efficiencies compared to those confined within mini flux ropes. Their dynamics are well characterized by prior analytical studies, including \cite{zharkova2004}, which describes analogous acceleration mechanisms in simplified current sheets. As such, we exclude these populations to focus on the helicity-dependent mini flux ropes, where electrons undergo efficient acceleration and are deflected toward distinct polarities, a process central to the novel findings of this work.

The asymmetric conjugate footpoints in spectral indices observed in our simulations are a ubiquitous feature in flare observations \citep{Jin2007, yang2012}. Previous studies attributed such asymmetries to the magnetic mirror effect \citep{aschwanden1999}, differing column densities from asymmetric reconnection \citep{emslie2003}, or anisotropic energy distributions \citep{mcclements2005}. In our simulations, however, the asymmetry originates from electron acceleration by distinct mini flux ropes, which transport nonthermal electrons into separate footpoints with different energy spectra. Notably, the resulting double sources are not magnetically connected. Their spatially correlated locations arise from neighboring mini flux ropes formed on adjacent regions of the current sheet, one of which develops opposite helicity due to tearing instability. While the spectral hardening in our simulation shows a break energy below the typical 100 keV threshold observed in previous observations \citep{kong2013, ning2019}, potentially due to resistivity uncertainties.

Additionally, several technical limitations warrant consideration. The resistivity, manually set as a hyperparameter in specific regions, complicates the integration of micro-scale particle effects into macro-scale electromagnetic fields. \added{In transitioning from the MHD simulation to the test particle approach, we apply an anomalous resistivity. This choice stems from the challenges in controlling the resistivity in the 3D grid-adaptive simulation, where the numerical resistivity varies spatially. We adopt an effective resistivity in the test particle study which is in line with a reasonable Lundquist number for the onset of tearing instability while mimicking the enhanced (kinetic-process dominated) resistivity for the particle simulation.} Moreover, the assumption of a fixed electromagnetic field could lead to unrealistic results, particularly in scenarios involving particle trapping. the dynamic evolution of mini flux ropes introduces time-dependent variability in electron precipitation signatures, complicating the interpretation of energy deposition patterns. \added{In our 3D MHD run, we simplify the energy equation by adopting an isothermal condition to reduce computational resource demands. By thereby achieve faster high-resolution simulation results with detailed fine structures. However, the thermal processes can play an important role in the formation of mini flux ropes and the development of turbulence \citep{wang2023,ye2023}, highlighting the need for more sophisticated approaches in future studies.} Higher-resolution MHD simulations may reveal more intricate processes within mini flux ropes. \added{The tearing and coalescence cascade of secondary flux ropes, which can be expected to occur at even higher Lundquist number regimes, could significantly differ from 2D simulations due to the inherently 3D topological nature of these structures, such as the variability of their cross-sections along the guide field direction}, suggesting the need for further refinement of our numerical approach. 

In conclusion, we present a comprehensive analysis of electron acceleration and transport within mini flux ropes in a large-scale current sheet beneath an erupting MFR. Our investigation of spatial and energy distributions of electron sources provides novel insights into the formation mechanism of conjugate double sources, consistent with observational results. The distinct acceleration processes within mini flux ropes of opposite magnetic helicity emerging from tearing instability, explain the observed broken power-law energy spectrum and asymmetric sources. These findings advance our understanding of particle dynamics within current sheets and provide a new framework for interpreting solar flare observations. Future work should include lower chromospheric layers, such that column depth effects can be considered and the identification of precipitation sources can be translated to actual HXR emission.

\begin{acknowledgements}
H.W., Y.G., Y.S., X.L.K., and M.D.D. were supported by the National Key R\&D Program of China (2022YFF0503004, 2022YFF0503002, 2021YFA1600504, and 2020YFC2201200) and NSFC (12333009). The numerical computation was conducted in the High Performance Computing Center (HPCC) at Nanjing University. RK acknowledges funding from KU Leuven C1 project C16/24/010 UnderRadioSun and the Research Foundation -- Flanders FWO project G0B9923N Helioskill.
\end{acknowledgements}

\clearpage

\bibliographystyle{plain}
\bibliography{ms}
\bibliographystyle{aasjournal}

\clearpage

\clearpage

\clearpage

\clearpage

\clearpage

\clearpage

\clearpage

\clearpage
\listofchanges
\end{document}